\documentclass[10pt,conference]{IEEEtran}
\IEEEoverridecommandlockouts
\usepackage{amsmath,amssymb,amsfonts}
\usepackage{algorithmic}
\usepackage{textcomp}
\usepackage[ruled,vlined]{algorithm2e}
\SetAlFnt{\small}
\usepackage{graphicx}
\usepackage{changepage}
\usepackage{xcolor}
\usepackage{biblatex}
\usepackage{comment}
\usepackage{amsthm}
\usepackage{adjustbox}
\usepackage{subfigure}

\theoremstyle{definition}

\def\BibTeX{{\rm B\kern-.05em{\sc i\kern-.025em b}\kern-.08em
    T\kern-.1667em\lower.7ex\hbox{E}\kern-.125emX}}

\begin{document}

\title{Quantum Communication in 6G Satellite Networks: Entanglement Distribution Across Changing Topologies}

\author{\IEEEauthorblockN{A. Sen, C. Sumnicht, S. Choudhuri, A. Chang, G. Xue}
\IEEEauthorblockA{\textit{School of Computing and Augmented Intelligence} \\
\textit{Arizona State University}\\
Tempe, AZ 85287, USA \\
\{asen, csumnich, s.choudhuri, ahchang, xue\}@asu.edu}
}

\maketitle

\begin{abstract}

As LEO/VLEO satellites offer many attractive features, such as low transmission delay, they are expected to be an integral part of 6G. Global {\em entanglement distribution} over LEO and VLEO satellites network must reckon with satellite movement over time. Current studies do not fully capture the dynamic nature of satellite constellations. We model a dynamic LEO/VLEO satellite network as a {\em time-varying graph} and construct a sequence of static graphs to represent a dynamic network. We study the entanglement distribution problem between a set of source-destination node pairs in this dynamic network utilizing {\em Multi-commodity Flow} (MCF). Solving MCF over a sequence of graphs independently for each graph may produce a completely different set of paths.  Changing the set of paths every time the graph topology changes may involve a significant amount of {\em overhead}, as an established set of paths must be taken down and a new set of paths established. We propose a technique that will avoid this overhead by {\em computing only one set of paths $\textit{P}$ to be used over all the graphs in the sequence}. The degraded performance offered by $\textit{P}$ may be viewed as the {\em cost} of using $\textit{P}$. The {\em benefit} of using $\textit{P}$ is the {\em overhead cost of path switching that can be avoided}. We provide a {\em cost-benefit analysis} in a LEO/VLEO constellation for entanglement distribution between multiple source-destination pairs. Our extensive experimentation shows that a significant amount of savings in overhead can be achieved if one is willing to accept a slightly degraded performance.

\end{abstract}


\section{Introduction}
Non-Terrestrial Networks (NTN) with Low-Earth-Orbit (LEO) and Very-Low-Earth-Orbit (VLEO) satellites will play a major role in the creation of 6G networks. In a recent paper, researchers from Huawei made a comprehensive discussion on their vision of VLEO-based NTN for 6G \cite{Huawei}. Recognizing the important role of satellites in the 6G environment, IEEE organized a workshop on {\em Advanced Solutions for 6G Satellite Systems} in 2022 \cite{IEEE6GWorkshop}. 
The European Space Agency has recently published a report on {\em 6G and Satellites: Intelligent Connectivity for a Sustainable Future} \cite{ESA}. Several academic studies on Satellite-6G network integration roadmap and coverage enhancement for 6G satellite-terrestrial integrated networks have also been recently published \cite{Deb,SC}.

\begin{figure}[tbh]
	\begin{center}
		\includegraphics[width = 0.5\textwidth, keepaspectratio]{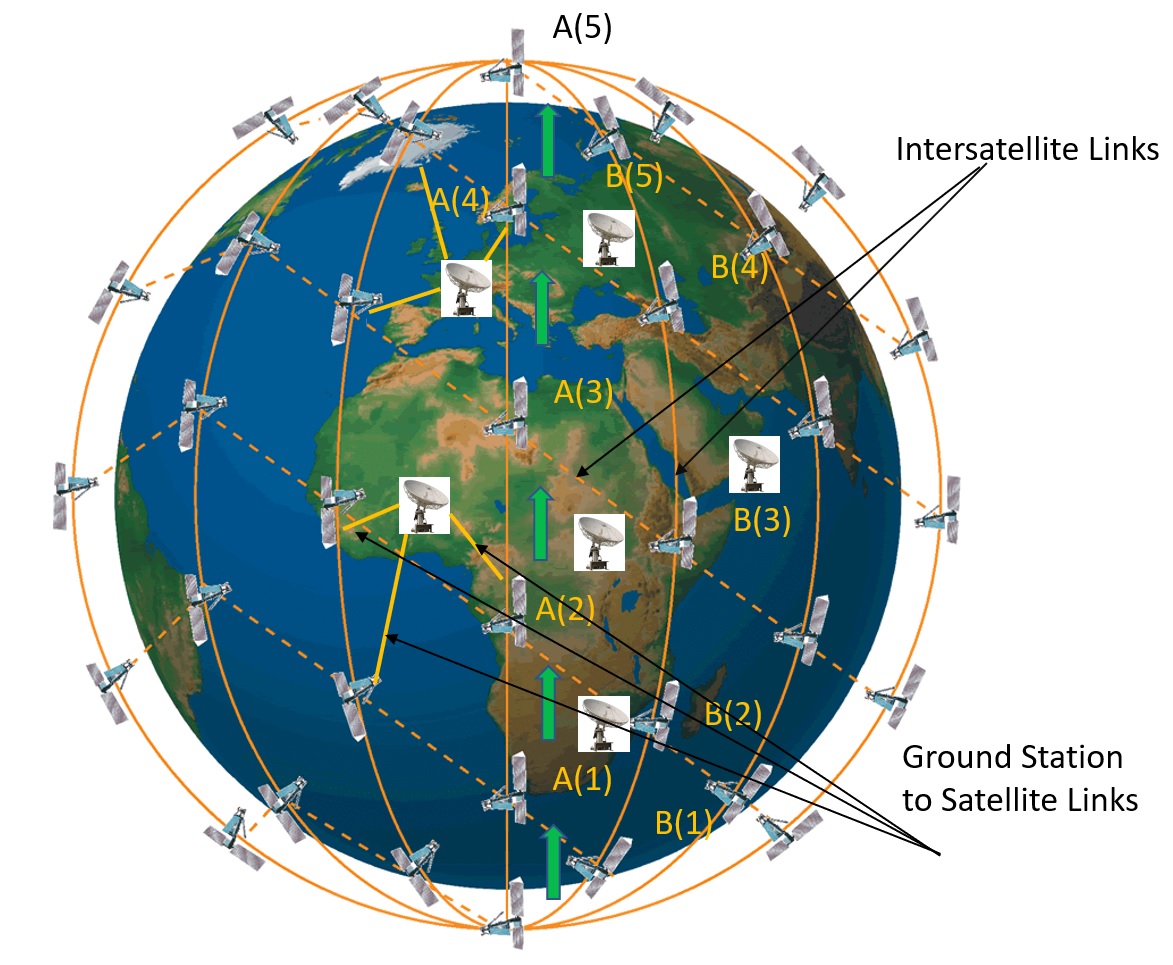}
        \caption{A Typical Satellite Constellation Network}
        \vspace{-4.0mm}
		\label{SatConstl}       
	\end{center}
\end{figure}

Just like 6G technology, Quantum Communication is also emerging as a future mode of long-distance communication. Quantum communication exploits the laws of quantum physics to guarantee privacy in a way classical communication cannot approximate, sparking significant interest among the research community and governments across the world in its development. The advantage quantum networks hold over classical networks mainly stems from the {\em spooky phenomenon} of {\em entanglement}, which has no classical analog~\cite{VanMeter2014Wiley}. 
An {\em ebit}, defined as a two-party entangled quantum state, comprises the fundamental unit of currency in quantum communications~\cite{Wikipedia}. Our work is concerned with the distribution of ebits.

Using local operations and classical communication (LOCC), repeaters in a quantum network can generate entanglement between adjacent repeaters. If a repeater Alice shares an entangled pair of photons (qubits) with another repeater, Charlie and Charlie shares an entangled photon pair with a repeater Bob, Charlie can perform entanglement swapping to induce Alice and Bob's qubits at their respective sites to collapse to an entangled state, in turn destroying the entanglements between Alice and Charlie and between Charlie and Bob. Repeaters can perform entanglement swapping to effectively extend entanglements until two remote parties, who may not be directly connected by an optical fiber or a free space optical link, share an entanglement. Using the resulting ebit, one party can teleport the state of a data qubit from their site to that of the other party without physically transmitting the data qubit itself, destroying the entanglement in the process.

Physical implementations of quantum networks have occupied both Earth~\cite{Alshowkan2021PRXQuantum} and space~\cite{Lu2022RevModPhys}.
The entanglement distribution rate over optical fibers is upper bounded by the repeaterless bound~\cite{Pirandola2017NatCommun} due to photon loss which scales exponentially with distance.
In contrast, the low altitude of LEO satellites as well as a wave of recent advances in space quantum communications as the most probable frontier for realizing global entanglement distribution~\cite{DeParny2023CommunPhys, Sidhu2021IETQuantumCommun}.

We study the entanglement distribution problem in alignment with \cite{Gundogan2021NpjQuantumInf}, which envisions both ground and space repeaters.
Both ground stations and LEO satellites can be equipped with 
(1) entanglement sources to generate entangled photon pairs,
(2) quantum non-demolition (QND) measurement devices to indicate whether an entanglement distribution attempt is successful, and
(3) quantum memories (QMs) to store photons.
QND and QM equipment is necessary for entanglement swapping.
Repeaters in this hybrid scheme can utilize {\em uplink, downlink} and {\em inter-satellite} links to distribute entanglement to any two users on Earth who request it.
The novelty of this scheme lies in the fact that it allows us to avoid terrestrial entanglement distribution altogether.

A LEO satellite in a constellation moves in a periodic manner.
The dynamic nature of such a network means photon transmission may be feasible between two parties at one point in time but infeasible at another time.
For example, if a satellite generates an entangled photon pair and wishes to transmit one photon to another satellite and store the other photon in a QM, there is no guarantee the satellite is close enough to its target satellite to support viable photon transmission throughout the entirety of a specified time interval.

Previous works studying satellite-based entanglement distribution from a networking perspective include~\cite{Khatri2021NpjQuantumInf, Panigrahy2022NetSciQCom, Chang2023arXiv}.
The problem studied in~\cite{Khatri2021NpjQuantumInf} considers satellite movement over time but exclusively distributes entanglement using the double downlink configuration, in which a satellite beams down an entangled photon pair to a ground station pair along two downlink channels.
By solely relying on the double downlink configuration, this model applies entanglement swapping strictly to ground repeaters and does not permit entanglement distribution along inter-satellite links. 
The problem studied in~\cite{Panigrahy2022NetSciQCom} only exercises the double downlink configuration and considers a satellite constellation at a fixed point in time, disregarding satellite movement over time.

The study conducted by Chang {\em et al.} in~\cite{Chang2023arXiv} is most likely the first one to take into account satellite movement over time and consider entanglement distribution along uplink, downlink, and inter-satellite links. It may be noted that due to the mobility of the satellites, the distance between them keeps changing, which in turn implies that links between two satellites and a satellite and a ground station may exist for some duration of time and may not exist for some other duration of time. 
If we construct a graph $G = (V, E)$ corresponding to the satellite constellation, where node $v \in V$ represents either a ground station or a satellite and an edge $e \in E$ represents a link between a ground station and a satellite or between two satellites, due to mobility of the satellites the graph $G = (V, E)$ will be a {\em dynamic} one as the edge set $E$ will keep changing with time. As the node set remains unchanged while the edge set keeps changing, this dynamic graph may be represented as $G(t) = (V, E(t))$. Since the movement pattern of the satellites is {\em periodic}, the {\em sequence} of graphs $G(1), G(2), \ldots$ repeats itself after a while, i.e., $G(1) = G(T + 1)$, $G(2) = G(T + 2)$, etc., where $T$ is the time period of repetition. This implies that we need to focus our attention only on a finite sequence of $T$ distinct graphs $G(1), G(2), \ldots, G(T)$.

The study conducted in \cite{Chang2023arXiv} does not consider a sequence of graphs $G(1), G(2), \ldots, G(T)$, but instead focuses on a specific graph $G(k), 1 \leq k \leq T$.  They consider a time interval $[\tau, \tau + \delta]$ during which the communication links (i.e., the edges of the graph) don't change. With this constraint, their model cannot capture the dynamic behavior of the network graph to its fullest extent as they focus only on one {\em static} graph $G(k)$. The work culminating in this paper marks a departure from the work of our predecessors in that we try to capture the true dynamic behavior of the graph $G(t)$, which implies that we consider entanglement distribution over a sequence of graphs $G(1), G(2), \ldots, G(T)$, instead of just one static graph. $G(k)$. To the best of our knowledge, this concept has never been explored in the context of  entanglement distribution over satellite networks.  

The contributions of this paper to the satellite-based entanglement distribution problem (described in Section 
\ref{sec2}) are as follows.

\begin{itemize}
    \item We study the entanglement distribution problem in a truly dynamic network of LEO satellites by modeling it as a time-varying graph $G(t) = (V, E(t))$
    \item The entanglement distribution problem is formulated as a variant of the {\em Multi-commodity Flow Problem (MCF)}, which is well studied in literature \cite{AMO}. However, unlike the classical MCF, which is solved for one graph, in the dynamic network environment, we need to solve MCF over a sequence of graphs $G(1), G(2), \ldots, G(T)$.
    \item Solving MCF over a sequence of graphs $G(j), 1 \leq j \leq T$, {\em independently} for each graph $G(j)$, may produce a completely different set of  paths ${\mathcal P}_{max}(j), 1 \leq j \leq T $, where ${\mathcal P}_{max}(j)$ maximizes the flow in graph 
    $G(j)$. Changing the set of paths every time the graph topology changes from $G(j)$ to $G(j+1)$ may involve a significant amount of {\em overhead}, as an established set of paths have to be taken down and a new set of paths have to be established. We present a technique that will avoid this overhead by computing {\em only one set of paths ${\mathcal P}$ to be used over all the graphs $G(j), 1 \leq j \leq T$}.  The set of paths ${\mathcal P}$ will have the following property: The {\em throughput} of the path set ${\mathcal P}$ may not be {\em optimal} in any graph $G(j), 1 \leq j \leq T$ (i.e., may not be equal to the  throughput of the path set ${\mathcal P}_{max}(j)$), but {\em close}  to optimal in every graph  $G(j), 1 \leq j \leq T$. Formally ``close'' to optimal is defined in the following way: Suppose that the path set ${\mathcal P}_{max}(j)$ finds the maximum throughput in graph $G(j)$ and $TP(\mathcal P (j))$ and  $TP({\mathcal P}_{max}(j))$ represent the throughputs of the path sets $\mathcal P$ in and ${\mathcal P}_{max}(j))$ in graph $G(j)$ for the set of requests $(s_i, t_i, d_i, 1 \leq i \leq k)$. The throughput of the path set $\mathcal P$ will have the property that $TP(\mathcal P(j)) \geq \epsilon \cdot TP(\mathcal P)_{max}(j)), \forall j, 1 \leq j \leq T$,
   where $\epsilon, 0 \leq \epsilon \leq 1$ is a constant. Our {\em entanglement maximization problem} aims to find the path set $\mathcal P$ that maximizes the value of $\epsilon$.
    \item The path set ${\mathcal P}$ will always provide somewhat degraded throughput in comparison with throughput produced by ${\mathcal P}_{max}(j), \forall j, 1 \leq j \leq T$. If the degraded performance is viewed as a {\em cost}, avoidance of the {\em overhead} associated with path switching may be viewed as a {\em benefit}. In Section \ref{sec4}, we present the result of our experimental evaluation of this cost-benefit study.
\end{itemize}


\section{Problem Formulation}
\label{sec2}

\begin{figure}[tbh]
	\begin{center}
		\includegraphics[width = 0.4\textwidth, keepaspectratio]{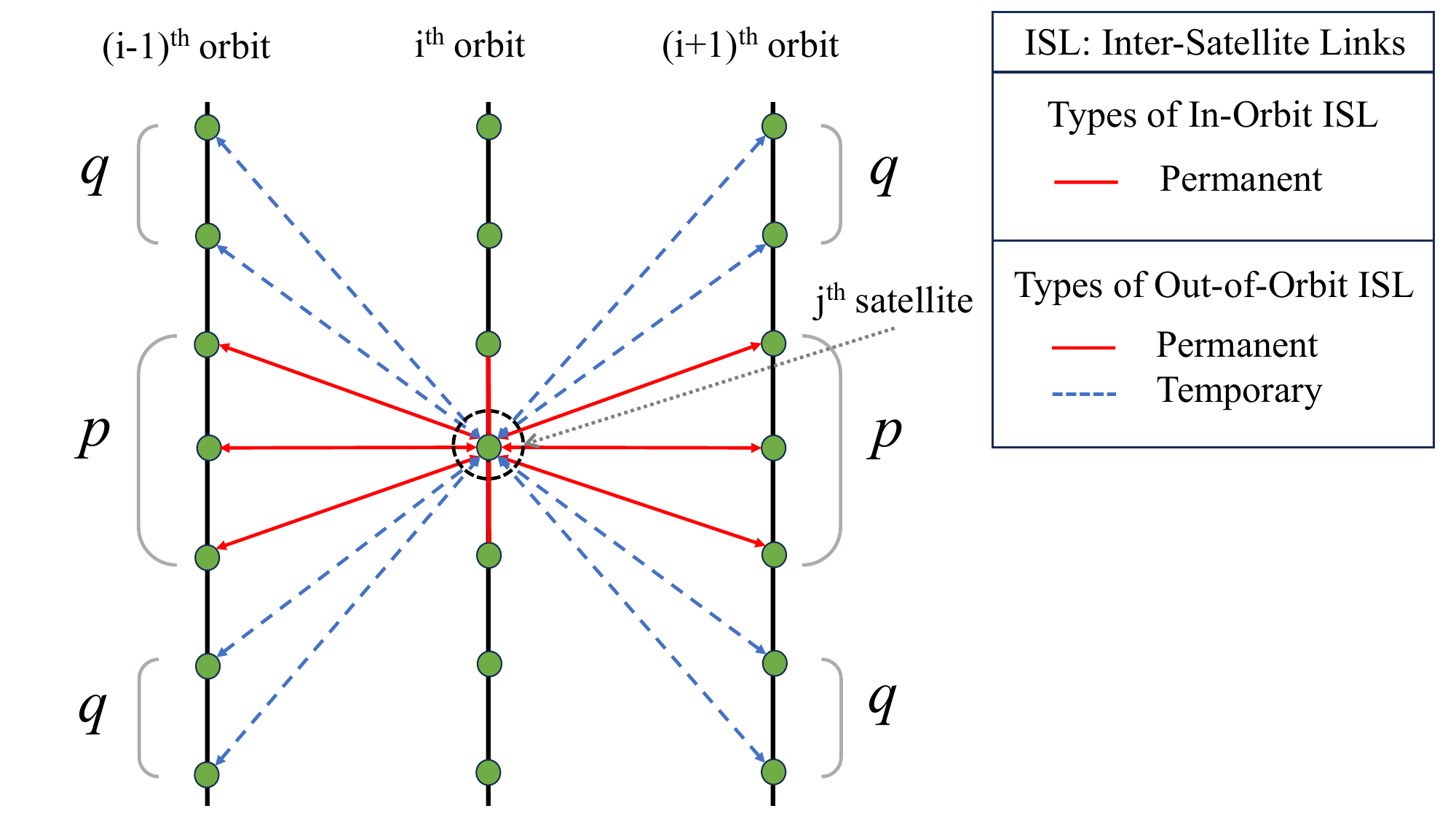}
        \caption{In-Orbit and Out-of-Orbit Inter-Satellite Links}
        \vspace{-4.0mm}
		\label{IntOrbConn}       
	\end{center}
\end{figure}

A typical satellite constellation, with satellites in multiple orbits, ground stations, inter-satellite links, and ground-station-to-satellite links, is shown in Fig. \ref{SatConstl}. As satellites continuously move in a fixed orbit, the distance between a pair of satellites keeps changing. The onboard transceivers on satellites and transceivers in ground stations have a fixed communication range. They can only communicate with each other if the distance between them does not exceed the communication range. Due to the movement of the satellites, inter-satellite links (ISL) may exist during some time intervals and may not exist during some other time intervals, resulting in a dynamic network. This dynamic network can be viewed as a sequence of static graphs $G(j), 1 \leq j \leq T$. The network topology evolves from $G(j)$ to 
$G(j+1), 1 \leq j \leq T - 1$, where $T$ represents the {\em periodicity} of the orbits (i.e., $G(1) = G(T+1)$). Our satellite network model comprises the following parameters:

\vspace{0.1in}
\noindent
{\bf System Model}
\begin{itemize}
    \item Number of orbits: $n$
    \item Number of satellites in each orbit: $m$
    \item Number of {\em permanent links} between satellites belonging to neighboring orbits: $p$ (ISL links that do not change over time are called permanent)
    \item Number of {\em temporary links} between satellites belonging to neighboring orbits: $2q$
    (ISL links that do change over time are called permanent)\\
    (Note: The Fig. \ref{IntOrbConn} shows satellites on three consecutive orbits $(i-1, i, i+1)$. The permanent and temporary ISL links of the $j-$th satellite on the $i-$th orbit are shown in red and blue, respectively, in Fig. \ref{IntOrbConn}. In this figure $p = 3$ and $q = 2$

    \item Number of {\em source-destination} pairs for entanglement distribution: $k$
    \item Number of trials used for each {\em source-destination} pair in experimentation: $r$
    \item Number of time periods (i.e., Periodicity, same as the number of graphs): $T$
    \item Bandwidth of a permanent link:  $\{1, \ldots, B_p\}$
    \item Bandwidth of a temporary link:  $\{0, \ldots, B_t, B_t < B_p\}$\\
     (Note: Bandwidth of both permanent and temporary links depends on distance and several other factors. Accordingly, we assume that the bandwidth of a permanent link can vary between 1 to an upper bound $B_p$, and similarly, the bandwidth of a temporary link can vary between 0 to an upper bound $B_t$. As the distance between two satellites having a permanent link will always be smaller than the distance between two satellites having a temporary link, we assume that $B_t < B_p$.)

    \end{itemize}

\vspace{0.1in}
\noindent
{\bf Assumptions/Constraints}

\noindent
(i) Each node $v_i \in V$ has a set transmitters $T_i$ and  a set receivers $R_i$.\\
(ii) {\em ``Capacity''} of an edge $e_{i,j}$, denoted as  $C(e_{i,j})$, is taken to be $minimum~ (|T_i|, |R_j|)$. Since the number of transmitters and receivers in a node $v_i$ are not required to be equal, $C(e_{i,j})$ may be different from $C(e_{j,i})$.  Thus from the capacity perspective, the orientation of an edge is important, and as such, we will view the graph $G= (V, E)$ as a {\em directed} graph.\\
(iii) We have a set of source-destination node pairs\\ $(s_1, t_1), (s_2, t_2). \ldots, (s_k, t_k)$ that needs entanglement exchange.\\
(iv) Each source-destination pair ($s_i, t_i)$ has {\em demand} value $d_i > 0$ associated with it, representing the number of ebits it wants to transmit\\
(iv) We need to establish a path $P_i$, corresponding to the source-destination node pair $(s_i, t_i), 1 \leq i \leq k$, through which entanglement exchange between $s_i$ and $t_i$ will take place.\\
(v) Entanglement exchange over a directed edge $e_{i, j}$ is limited by the capacity of that edge $C(e_{i, j})$.

\vspace{0.1in}
\noindent
{\bf Objective:} Find one set of paths ${\mathcal P}$ to be used in all the graphs $G(j), 1 \leq j \leq T$ that maximizes $\epsilon, 0 \leq \epsilon \leq 1$, under the constraint that $TP({\mathcal P(j)}) \geq \epsilon \cdot TP({{\mathcal P}_{max}}(j)), \forall j, 1 \leq j \leq T$, where $TP(\mathcal P(j))$ represents the throughput of the path set ${\mathcal P}$ in graph $G(j)$ and $TP({{\mathcal P}_{max}}(j))$  represents the {\em maximum throughput} possible for the set of requests $(s_i, t_i, d_i, 1 \leq i \leq k)$  in the graph $G(j)$. It may be noted that although the path set ${\mathcal P}$ doesn't change in different graphs $G(j)$, as the bandwidth of the permanent links changes in different graphs, the throughput of the path $\mathcal P$ in graph $G(j)$ may be different from the throughput of $\mathcal P$ in graph $G(j+1)$.

\section{Entanglement Distribution Path Computation}
\label{sec3}

We built upon Multi-commodity Flow (MCF) \cite{AMO} to solve the Entanglement Distribution Paths Computation problem. The novelty of our solution is that unlike the classical MCF problem, which is solved over only one graph, we need to solve it over a sequence of graph $G(j), 1 \leq j \leq T$. As our solution is built upon classical MCF, we provide a short discussion next.

\subsection{Multi-commodity Flow Over a Single Graph}
\label{CMCF}

Given a flow network $G = ( V , E )$ where edge $( u , v ) \in E$  capacity $c ( u, v )$. There are $k$ commodities $K_1 , K_2 , \cdots , K _k$ defined by $K_i = ( s_i , t_i , d_i )$ where $s_i$ and $t_i$ are the {\em source} and {\em sink} of commodity $i$ and $d_i$ its {\em demand}. The variable $f_i ( u, v )$ defines the part of flow $i$ along edge $( u, v )$, where $f_i ( u, v ) \in [ 0, d_i ]$ in case the flow can be split among multiple paths, and $f_i ( u, v ) \in  \{ 0, d_ \}$ otherwise (i.e., "single path routing"). The goal of the MCF problem is to find an assignment of all flow variables which maximizes total flow subject to all the constraints:\\

{\em Objective:} {\em Maximize} $\sum_{i = 1}^k F_i$, where $F_i$ is the amount of $s_i$ to $t_i$ flow. \\

\begin{table*}[h!]
\vspace{0.3cm}
\begin{tabular}{|c||c|c|c|c|c|c|c|c|c|c|}
\hline
& \multicolumn{2}{c|}{T = 4} & \multicolumn{2}{c|}{T = 6} & \multicolumn{2}{c|}{T = 8} & \multicolumn{2}{c|}{T = 10} & \multicolumn{2}{c|}{T = 12} \\ \cline{2-11}
k & \shortstack{\% Drop in\\Throughput} & \shortstack{Cost of\\Switching} & \shortstack{\% Drop in\\Throughput} & \shortstack{Cost of\\Switching} & \shortstack{\% Drop in\\Throughput} & \shortstack{Cost of\\Switching} & \shortstack{\% Drop in\\Throughput} & \shortstack{Cost of\\Switching} & \shortstack{\% Drop in\\Throughput} & \shortstack{Cost of\\Switching} \\ \hline
3 & 20.64 & 268.40 & 24.72 & 425.20 & 28.56 & 590.60 & 30.00 & 767.40 & 30.27 & 933.60 \\ \hline
5 & 19.09 & 380.40 & 20.95 & 622.00 & 23.15 & 877.20 & 23.97 & 1151.40 & 25.48 & 1406.80 \\ \hline
7 & 25.06 & 420.40 & 27.37 & 732.40 & 29.95 & 1020.40 & 30.75 & 1311.20 & 33.68 & 1608.20 \\ \hline
9 & 22.97 & 425.80 & 25.42 & 735.00 & 28.94 & 1040.80 & 30.58 & 1348.40 & 33.21 & 1675.80 \\ \hline
\end{tabular}
\vspace{0.1cm}
\caption{Cost (\% drop in throughput) vs. Benefit (savings in path switching) over multiple sets of source-destination pairs ($k$) and multiple sets of graphs ($T$)}
\label{Cost-Benefit-Table}
\end{table*}

\noindent
(2)  All $F_i$ values must be limited by their respective demand values $d_i$
\[\forall i,  1 \leq i \leq k, F_i \leq d_i\]
(3) Link capacity: The sum of all flows routed over a link cannot exceed its capacity
\[\forall ( u , v ) \in  E, ~~~\sum {F_i}_{i = 1}^k ( u , v ) \leq  c( u , v )\]
(4) Flow conservation at the source node: $F_i$ is the difference between the outgoing and incoming $i-$th flow at the source node $s_i$
\[\forall i, 1 \leq i \leq k, \sum_{w \in N(s_i)} f_i ( s_i , w ) - \sum_{w \in {\bar N(s_i)}} f_i( w , s_i ) = F_i,\] where $N(u)$ is the set of all nodes $v\in V$ that has a directed edge from $u$ to $v$, and where ${\bar N(u)}$ is the set of all nodes $v\in V$ that has a directed edge from $v$ to $u$

\vspace{0.1in}
\noindent
(5) Flow conservation at the sink node: $F_i$ is the difference between the incoming and outgoing $i-$th flow at the sink node $t_i$
\[\forall i, 1 \leq i \leq k, \sum_{w \in {\bar N(t_i)}} f_i ( w , t_i ) - \sum_{w \in N(t_i)} f_i( t_i , w ) = F_i\]

\vspace{0.1in}
\noindent
(6) Flow conservation on transit (nodes other than the source and the sink) nodes: The amount of a flow entering and exiting a transit node $u$ must be equal


\[\forall i, 1 \leq i \leq k,  \sum_{w \in {\bar N(u)}} f_i( u , w ) - \sum_{w \in {N(u)}} f_i ( w , u ) = 0\] when $u \neq s_i, t_i$ 


\subsection{Multi-commodity Flow Over Multiple Graphs}

      We have $T$ graphs $G(1), \ldots, G(T)$, ($G(j), 1 \leq j \leq T$). In the classical Multi-Commodity Flow (MCF) problem, discussed in section \ref{CMCF}, we used the variable $f_i ( u, v )$ to denote the part of the $i-$the flow along the edge $( u, v )$.  
      Since now we are trying to compute MCF across multiple graphs $G(1), \ldots, G(T)$, 
      we introduce an additional index $j$ to the variable $f_i ( u , v )$ to keep track of the graph (among $G(1), \ldots, G(T))$, 
      that we are referring to. The modified variable $f_i^j ( u, v )$ denotes the part of flow $i-$th flow on the $j-$th graph along the edge $( u, v )$. Due to this modification, the variables and constraints stated for MCF in subsection \ref{CMCF} have to be appropriately modified.\\


\noindent
(7)  The total $i-$th flow across all the graphs $G(1), \ldots, G(T)$, must not exceed the demand $d_i$
\[\forall i,  1 \leq i \leq k, \sum_{j = 1}^T F_i^j \leq d_i\]
(8) Link capacity: The sum of all the flows routed over a link $(u, v)$ in the graph $G(j), 1 \leq j \leq T$, cannot exceed the capacity of the link $(u, v)$ in $G(j)$
\[\forall j \in 1 \leq j \leq T, \forall ( u , v ) \in  E(j) , ~~~\sum {{f_i}^j}_{i = 1}^k ( u , v ) \leq  c^j( u , v )\]
(9) Flow conservation at the source node: $F_i^j$ is the difference between the outgoing and incoming $i-$th flow at the source node $s_i$ in the $j-$th graph $G(j)$., i.e., 
$\forall i,  1 \leq i \leq k$ and $\forall j,  1 \leq j \leq T$ \[\sum_{w \in N(s_i)} f_i^j ( s_i , w ) - \sum_{w \in {\bar N(s_i)}} f_i^j( w , s_i ) = F_i^j,\] where $N(u)$ is the set of all nodes $v\in V$ that has a directed edge from $u$ to $v$, and where ${\bar N(u)}$ is the set of all nodes $v\in V$ that has a directed edge from $v$ to $u$



\vspace{0.1in}
\noindent
(10) Flow conservation at the sink node: $F_i^j$ is the difference between the incoming and outgoing $i-$th flow at the sink node $t_i$ in the $j-$th graph $G(j)$, i.e., 
$\forall i, 1 \leq i \leq k$ and $\forall j, 1 \leq j \leq T$ \[\sum_{w \in {\bar N(t_i)}} f_i^j ( w , t_i ) - \sum_{w \in N(t_i)} f_i^j( t_i , w ) = F_i^j\]



\noindent
(11) Flow conservation on transit (nodes other than the source and the sink) nodes: The amount of a flow entering and exiting a transit node $u$ must be equal in all the graphs  $G(1), \ldots, G(T)$, i.e., 
$\forall i, 1 \leq i \leq k$ and $\forall j, 1 \leq j \leq T$             
\[\sum_{w \in {\bar N(u)}} f_i^j( u , w ) - \sum_{w \in {N(u)}} f_i^j ( w , u ) = 0\] when $u \neq s_i, t_i$  

\vspace{0.1in}
\noindent
(12) As we would like to avoid path switching (i.e., to maintain the same set of paths over all the graphs, $G(1), \ldots, G(T)$), we need the constraint that for any two graphs $G(j_1)$ and $G(j_2), 1 \leq j_1, j_2 \leq T, j_1 \neq j_2$ and $\forall i, 1 \leq i \leq k$, \[ (f_i^{j_1}( u, v ) > 0) \iff  (f_i^{j_2}( u, v ) > 0)\]  

This logical constraint can be incorporated in the ILP with the introduction of a new binary variable $\delta$ and the following set of four constraints ($M$ is a large integer):
\[f_i^{j_1}( u , v ) \geq 0.000001 - M \cdot (1 - \delta)\]
\[f_i^{j_1}( u , v ) \leq M \cdot \delta\]
\[f_i^{j_2}( u , v ) \geq 0.000001 - M \cdot (1 - \delta)\]
\[f_i^{j_2}( u , v ) \leq M \cdot \delta\]


{\em Achieved Throughput} of a set of $k$ paths in a graph $G(j)$: 
\[\sum_{i = 1}^k F_i^j\]


{\em Maximum Throughput} of a set of $k$ paths in a graph $G(j)$ 
is denoted by \(MaxTP(j, k)\).

It may be noted that the computation of \(MaxTP(j, k)\) is the same as the computation of MCF in the graph $G(j)$, which can be done using well-known techniques \cite{AMO}. We define the parameter $r_j$ for the graph $G(j)$ as the {ratio} between the {\em Achieved Throughput} to the {\em Maximum Throughput} in $G(j)$.

\vspace{-0.1cm}

    \[r_{j} = \frac {\sum_{i = 1}^k F_i^j}  {MaxTP(j, k)}\]



    \[r_{min} = {\tt minimum_{1 \leq j \leq T}}~{r_j} \]

Our objective is to set up the flows in a manner that {\em may not achieve maximum throughput in any one of the graphs $G(1), \ldots, G(T)$}, but {\em will achieve throughput that will be close to the maximum throughput in every one of the graphs $G(1), \ldots, G(T)$}. This objective can be expressed in a formal way for use in our Integer Liner Program in the following way:
\[{\tt Maximize}~ \epsilon\] subject to the constraints:
\[\forall j, 1 \leq j \leq T, {\sum_{i = 1}^k F_i^j} \geq \epsilon \cdot  (MaxTP(j, k))\]

\section {Experimental Evaluation}
The parameters used in our system model (presented in Section \ref{sec2}) for experimental evaluation of our proposed technique are as follows: 
$n = 5, m = 5, p = 1, q = 1, B_p= 3, B_t =2$ In our experiments, the following values $k$ (the number of source-destination pairs) were used: 3, 5, 7, 9.  We conducted 5 trials for each value of $k$ (i.e., r = 5). The values for periodicity $T$ used are the following: 4, 6, 8, 10, 12. 

As indicated earlier, using the path set $\mathcal{P}$ across all the graphs $G(j)$ may result in degraded performance. However, the overhead cost of path switching from one graph to the next one can be avoided. In this paper, we measure the overhead cost of path switching from $G(j)$ to $G(j+1)$ in the following way. Suppose that path sets $\mathcal{P}_{max}^j$ and $\mathcal{P}_{max}^{j+1}$ maximize throughputs in the graphs $G(j)$ and $G(j+1)$ respectively 
and $E(\mathcal{P}_{max}^j)$ and $E(\mathcal{P}_{max}^{j+1})$ represent the edge sets used by the path sets $\mathcal{P}_{max}^j$ and $\mathcal{P}_{max}^{j+1}$ respectively. We take the Exclusive-OR of the edge sets $E(\mathcal{P}_{max}^j)$ and $E(\mathcal{P}_{max}^{j+1})$, denoted by $E_{XOR}^{j, j+1}$ and take it's cardinality as the measure of the cost of switching from the path set $\mathcal{P}_{max}^j$ in graph $G(j)$ to the path set  $\mathcal{P}_{max}^{j+1}$ in graph $G(j+1)$. The rationale for measuring cost in this way is guided by the fact that this is the set of edges that has to be {\em activated} or {\em deactivated} in transitioning from the path set $\mathcal{P}_{max}^j$ in $G(j)$ to the path set  $\mathcal{P}_{max}^{j+1}$ in $G(j+1)$ (i.e., updates in the routing table at the satellite nodes). The {\em Total Cost} of path switching over the graph sequence  $G_1, \ldots, G_T$ is given by \(\sum_{j = 1}^{T-1} |E_{XOR}^{j, j+1}|\).

\begin{figure}[tbh]
	\begin{center}
		\includegraphics[width = 0.4\textwidth, keepaspectratio]{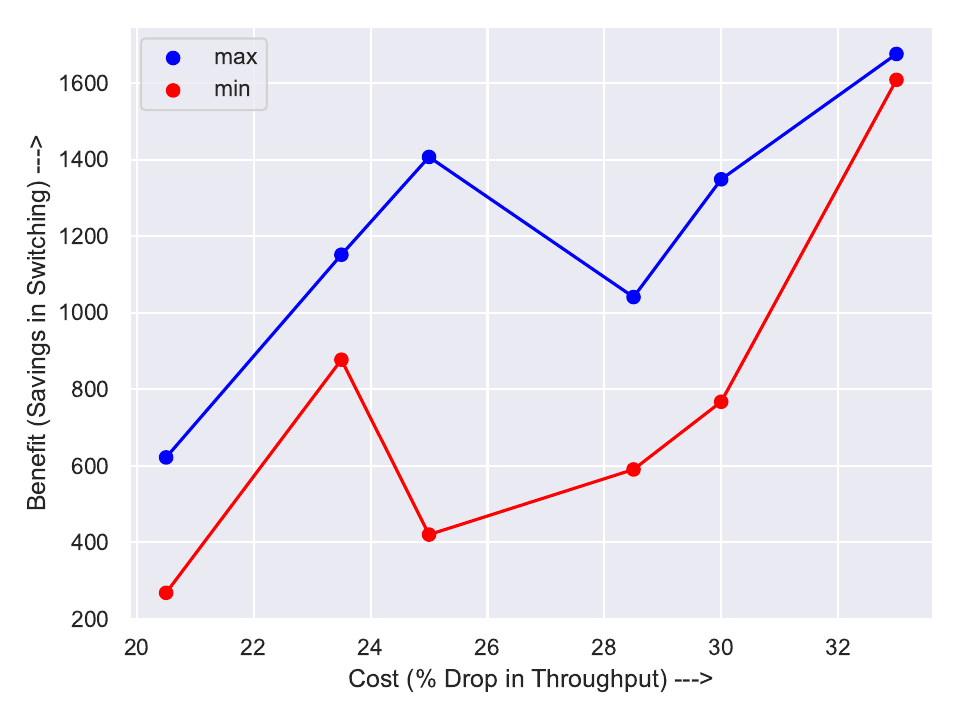}
        \caption{Cost (\% Drop in Throughput) vs. Benefit (Saving in Path Switching) Plot}
        \vspace{-9.0mm}
		\label{Cost-benefit}       
	\end{center}
\end{figure}

\begin{figure}[tbh]
	\begin{center}
		\includegraphics[width = 0.4\textwidth, keepaspectratio]{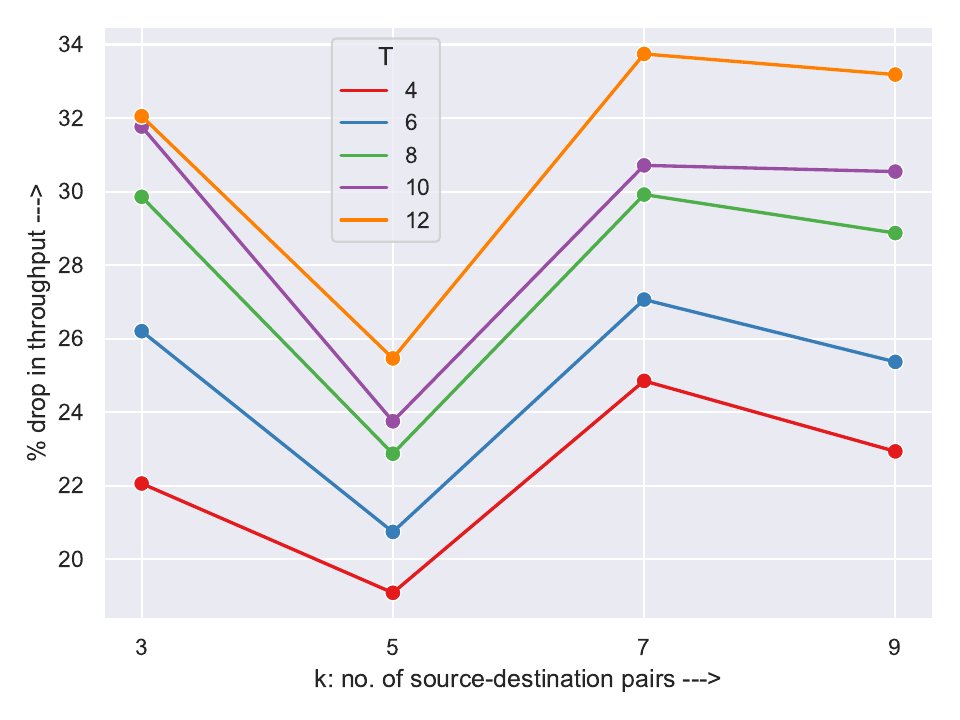}
	    \vspace{-4.0mm}
        \caption{Impact of the number of source-destination pairs on \% Drop in throughput}
        \vspace{-6.0mm}
		\label{kVsThroughputDrop}       
	\end{center}
\end{figure}

The optimal solution to each problem instance was computed using Gurobi 10.0.1. Table \ref{Cost-Benefit-Table} presents the evaluation results of the proposed technique over multiple sets of source-destination pairs and graphs. For a specific value  of $k$ (the number of source-destination pairs) experiment was conducted 5 times (i.e., r = 5). The average drop of throughput over 5 trials is presented in Table \ref{Cost-Benefit-Table}. It may be observed from the table that the smallest and largest drop in throughput was 19.09\% and 33.68\%, respectively. Similarly, the smallest and largest cost was 268.40 and 1675, respectively. From this table, one can infer that substantial savings can be made in terms of path-switching cost if one is willing to accept a certain amount of loss in throughput. As a typical example, from $k = 9$ and $T = 8$, 1040.80 units of cost savings can be made if one is willing to accept a 28.94\% drop in throughput. The cost-benefit trade-off is presented in Fig. \ref{Cost-benefit}. It shows that if one is willing to accept a 30\% drop in throughput, cost savings will be at least 800 and at most 1350. The impact of the number of source-destination pairs on the percentage drop in throughput is shown in Fig. \ref{kVsThroughputDrop}. From the plot, one can infer that the drop in throughput doesn't directly relate to the number of source-destination pairs, as it can be seen that drop in throughput decreased when $k$ increased from 3 to 5 and 7 to 9, and it increased when $k$ increased from 5 to 7. The actual choice of source-destination pairs seems to impact the drop in throughput more than the number of source-destination pairs.

\vspace{-0.3cm}

\vspace{-0.3cm}

\label{sec4}
\section {Conclusion}
\label{sec5}


This paper proposes a novel routing technique for entanglement distribution in a 6G satellite network. Our approach addresses the challenge of adapting to changing network topologies caused by satellite movements in orbits without requiring path modifications. While our routing scheme may exhibit slightly reduced performance compared to optimal routing, it presents notable cost savings associated with path-switching. Our comprehensive cost-benefit analysis demonstrates that a significant overhead reduction can be achieved by accepting a slight performance degradation.


\printbibliography
\end{document}